
\magnification=\magstep1
\baselineskip  22 true   pt
\hsize 5 in\hoffset=.4 true in
\vsize 6.9 in\voffset=.4  true  in

\def\a{\alpha}

\def\g{\gamma}
\def\d{\delta}
\def\e{\epsilon}

\def\l{\lambda}
\def\om{\omega}
\def\p{\pi}
\def\m{\mu}
\def\n{\nu}
\def\r{\rho}
\def\s{\sigma}

\def\QNE{{F^a_{\mu\n}}}
\def\QNF{{F^a_{\mu 0}}}
\def\QNEB{{{\tilde{F}}^a_{\mu\n}}}
\def\QNEBS{{{\tilde{F}}_{\mu\n}}}

\def\QNK{{1\over{g_5^2}}}

\def\QNII{{{\bar\eta}_{a\mu\n}}}

\def\lnrt{\longrightarrow}
\def\boxpi{{{\partial^2 \Pi}\over \Pi}}
\def\pip{{{\Pi '}\over \Pi}}
\def\prm{\partial_M}
\def\prl{\partial_L}
\def\tilo{{{{\tilde{\om}}_{\m}}}}
\def\xmhat{{{{\hat{x}}_{\m}}}}

\vfil
\headline{\hfill IP/BBSR/92-27.}
\vfil
\centerline {\bf{Charged Topological Solitons in 4+1 Dimensional YMCS Theory}}
\vskip 3.0 cm
\centerline {\bf{C.S.Aulakh}}
\centerline{ Institute of Physics, Sachivalya Marg}
\centerline{ Bhubaneshwar,751005,India}
\vskip 2.0 cm
\centerline{\bf{ ABSTRACT}}

\noindent
 We show that when a Chern-Simons term is added to the action of
 $SU(N)$ ($N\geq 3$) Yang-Mills theory in 5 dimensions the usual
self-dual topological solitons present in the theory necessarily pick
up a (topological) electric charge .

\vfil
\footline {{Email:aulakh\%iopb{@}shakti.ernet.in} {\hskip 2 cm}\hfil
{\hbox{{March 1992}}}}

\eject

In a recent paper ${}^1$ the present author suggested that an
examination of the properties of a certain class of co-winding
solitons (``syncyclons") generically present in higher dimensional
field theories (i.e defined on space-time extended beyond 4
dimensions by a compact space) is called for. Such solitons would
appear as point, string or sheet vacuum defects to our three
dimensional low energy eyes and might thus signal the presence of
otherwise unobservable extra dimensions.

The simplest paradigmatic example of such solitons is furnished by
Yang-Mills (YM) theory on $M^{4}\times S^{1}$. The basic solution is
nothing but the periodic instanton or ${\it caloron}^3$
reinterpreted as a soliton on $M^{4}\times S^{1}$. Classically, such
solitons have no (non-abelian) electric charge and, in the
Kaluza-Klein (KK) limit $r>>R$ (where R is the radius of $S^{1}$ and
$r$ the 3-dimensional distance),
their fields are those of a non-abelian magnetic dipole together
with a scalar potential coming from the extra compoment of the
vector potential${}^{1,4}$. However, in order to sketch a believable
picture of their properties it is necessary to take into account the
effects of coupling to fermions and the quantization of global gauge
collective coordinates since these can radically change the quantum
numbers labelling the soliton states due to fermion number
fractionalization${}^5$ and zero point (quantum) rotation in the
internal non-abelian  space. In particular using standard
results on fermion number fractionalization in odd dimensions${}^5$
one finds that the soliton picks up an abelian charge
{\hbox {$-{{\nu q}\over 4}$}}
 ($\nu$ the Pontryagin index) from each fermion of abelian charge
q (Baryon/Lepton number, Electric charge. etc). In addition the
expectation value for the charges corresponding to the $U(1)$s in the
decomposition {\hbox{$SU(N) \supset SU(2)\times U(1)^{N-2}$}}
is also proportional to $\nu/4$ (with the soliton trivially embedded
in the SU(2) factor). Thus if the gauge group is SU(3) then
$<Q^{8}>={-N_{f}\nu\over {4 {\sqrt 3}}}$
 where $N_{f}$ is the number of fermion
flavours. The term in the gauge effective action which represents
this effect is nothing but the 5 dimensional Chern Simons term times
$N_{f}$. This term is the low-momentum approximation of the fermion
determinant in the gauge field background. In order to determine
the gauge quantum numbers of the soliton one must quantize the
collective coordinates of a static solution to the field equations
of the YM-Chern  Simons (YMCS) system in 5 dimensions. This problem
bears a close analogy to that of determining the quantum numbers of
the Skyrmeon${}^7$. The roles of flavor and color are interchanged and
the Chern-Simons term is the analog of the WZW term which also arises
by integrating out fermions coupled to the Skyrme field . The
significant generalization is that the present system is gauge
invariant while the Skyrme-WZW Lagrangian has only a global (flavour)
invariance. Since the induced charge described above is coupled to a
gauge field it follows that here the static solution one must
quantize around is a ${\it charged}$ soliton. In the Skyrme case the
effect of the WZW term in the field equations is to produce a
deformation of the original soliton without obviously changing its
external interactions (as opposed to its flavour and spin quantum
numbers). On the contrary, in the present case, the as we shall show,
the deformed solution carries a (topological) electric charge.
 Charged solitons arise in YMCS-Higgs systems in 2+1
dimensions as well${}^8$ where the presence of the CS term allows
charged vortex solutions to exist.
In 2+1 dimensions the CS term makes the gauge particles massive so that
the induced charge couples to a short range force carrier. The
integral of the Higgs field charge density is proportional to the
quantized flux (divided by 4) so that there too the electric charge
is fractional and topological in nature.
 The existence of charged vortex solutions
in 2+1 dimensions can be seen as the analog of the ``Witten effect''
in 4 dimensions whereby the addition of a CP violating $\theta$ term
(i.e a Pontryagin density) to the YM-Higgs lagrangian results in
charged monopole solutions${}^6$.  One therefore expects that
such a conversion of ``magnetic" charge neutral solitons to charged
ones occurs in all odd dimensions provided the theory without the CS
term has a topological soliton solution. Note however
\vfil
\eject

\noindent that in the present case the magnetic
fields are asymptotically {\it{dipolar}}.

As a preliminary to the quantization${}^9$
of the collective coordinates of solitons in YM theory on
$M^{4}\times S^{1}$, and also because of their intrinsic interest
we have studied static solutions of the YMCS system in $M^{5}$. The
generalization of our results to $M^{4}\times S^{1}$ and the
situation in higher odd dimensions will be reported separately${}^{10}$.
 We will show that the above heuristic argument for the existence
of charged topological solitons on 5 dimensions is borne out by a
detailed analysis of the SU(N), YMCS action:
$$\eqalign {  S &= \int d^5 x ({1\over {2 g_5^2}} tr {F_{MN}}^2 +
 {{i N_f}\over {48\pi^2}} \omega_5) \cr
 \omega_5 &=\e_{MNLPQ}\quad tr(\prm A_N(\prl A_P A_Q + {3\over 2} A_L A_P A_Q)
 + {3\over 5} A_M A_N A_L A_P A_Q )}
\eqno(1)$$

The field equations are:

$$ D_M F_{MN}^A = -{{N_f g_5^2}\over {128\pi^2}}\e_{NMLPQ}\quad tr(\l^A
F_{ML} F_{PQ}) \eqno(2) $$

Our normalizations and definitions are

$$\eqalign{ A_M &= A_M^A {{\l^A}\over 2i} \qquad\qquad tr \l^A\l^B =2\d^{AB}\cr
F_{MN} &=F_{MN}^A {{\l^A}\over 2i}=\partial_{[M} A_{N]} + [A_M,A_N]\cr
A,B,...&= 1,...,N^2-1 \cr
M,N,...&= 0,1,2,3,4\cr
\m,\n,...&= 1,2,3,4 \cr
a,b,...&=1,2,3\cr
{\hat a},{\hat b},...&=4,...,N^2-1}\eqno(3)$$

Here $\{\lambda^A\}$ are the Gell-Mann matrices for SU(N). Our metric
convention is $(-++++)$.
In the absence of the CS term ($N_f=0$) we have the usual self dual
solutions${}^{11}$  in the SU(2) Euclidean sector:

$$\eqalign{ A_0&=\partial_0 =0\cr
A_{\m} &=-\QNII {{\tau^a}\over 2i} \partial_{\n} ln\Pi =
i{{{\overline\Sigma}_{\m \n}}}\partial_{\n} ln\Pi \cr
\Pi^{-1}\partial^2\Pi &=0 \qquad\qquad \Pi =1
+\sum_{i=1}^K {{\r_i^2}\over {(x-x_i)^2}} \cr
\n &= {1\over{32\p^2}}\int d^4x \QNE \QNEB = K }\eqno(4)$$

where $\r_i,x_i$ are the scale factors and positions of the K solitons and
{\hbox{$\n=K$}}  the total winding number.
 In the above $\{\QNII\}$,$\tau_a,\Pi$
are the anti-self-dual 't Hooft symbols, Pauli matrices and ``super potential"
respectively. Fields outside the $SU(2)$ subalgebra  generated by
$\{\lambda^{a}\}$ play no role. The asymptotics are those of a
non-abelian magnetic dipole with zero electric charge (since $F_{\mu
0}=0)$. The field energy is

$${\cal E} = {\QNK} \int d^4x({1\over 4} \QNE^2 + {1\over 2} \QNF^2)
= {{8\p^2\n}\over{g_5^2}}\eqno(5)$$

We now show that these solutions become charged when the effect
 of the C.S. term is included. For simplicity let us work with $\n=K = 1$
 and $ N = 3$. Generalizations are obvious and will be indicated below.
  The field equations are

$$ D_{M} F_{M\n}^A = {{N_f g_5^2}\over {16\pi^2}}\quad
 tr(\l^A F_{\m 0}\QNEBS) \eqno(6a)$$

$$ D_{\m} F_{\m 0}^A = -{{N_f g_5^2}\over {64\pi^2}}
 tr(\l^A F_{\m \n}\QNEBS) \eqno(6b) $$

It is easy to check that for $A = 8$ the ansatz of (4) gives zero for
the l.h.s of eqn.(6b) while the rhs is proportional to
the Pontryagin density and is hence nonzero.
 Thus (4) is no longer an adequate ansatz when $N_f\neq 0$.
The required modification is obviously:

$$\eqalign{A_{\m}^A &=0\qquad\qquad A=4,5,6,7,8\cr
A_{\m}^a &=-\QNII  \partial_{\n} ln\Pi \qquad\qquad \Pi=\Pi(x^2)\cr
A_0^A &=0=\partial_0 \qquad A\neq 8} \eqno(7)$$

Spherical symmetry ensures that the r.h.s of the equation for
${D_{\m}F_{\m 0}^a}$ vanishes as it must for consistency. The only
nontrivial equations are then :
$$\eqalign {{D_{\m}F_{\m\n}^a} &=
-\a \partial_{\m} A_0^8 {{{\tilde{ F}}_{\m \n}^a}} \cr
 \partial^2 A_0^8 &= {\a \over 4}\QNE \QNEB
 = {\a \over 2}\partial_{\mu}{{{\tilde{\om}}_{\m}}}\cr
 \a&={{N_f g_5^2}\over {{32\pi^2}{\sqrt {3}}}}\cr
 {{{\tilde{\om}}_{\m}}}&= \e_{\m\n\l\g}(A_{\n}^a\partial_{\l}A_{\g}^a +
 {{\e_{abc}}\over 3} A_{\n}^aA_{\l}^b A_{\g}^c)}\eqno(8)$$

Thus the Pontryagin density serves as the charge density in a Poisson
equation for the electrostatic potential in the 8 direction. The
solution for $A_0^8$ is simply

$$A_0^8(x) = -{\a\over{16\pi^2}}\int d^4y{1\over{(x-y)^2}} \QNE(y) \QNEB(y)
\eqno(9) $$
and also

$$ \partial_{\m} A_0^8 = {\a \over 2}({{{\tilde{\om}}_{\m}}} +
 {{D {{{\hat{x}}_{\m}}}\over {x^3}}}) \eqno(10) $$

The arbitrarinesss represented by D is crucial to the existence of a
charged solution. It is easy to check
(assuming that the deformed superpotential
has the same leading behaviours as $x\lnrt 0$ and $x\lnrt\infty$ as before)
that the leading behaviour of
$\tilo$  as $x\lnrt\infty\quad$  is $O(x^{-7})$ . Thus to obtain the
electric field corresponding to the charge $Q^8=-{N_f\over{4\sqrt 3}}$
we must choose $D=8$.

The other equation is now

$$ D_{\m} F_{\m\n}^a  =-{{\a^2} \over 2}({{{\tilde{\om}}_{\m}}} +
 {{D {{{\hat{x}}_{\m}}}\over {x^3}}}) {{{ \tilde{F}}_{\m \n}^a}}\eqno(11) $$

Due to spherical symmetry the above equation reduces to

 $$\eqalign{(\boxpi)' -2(\pip)(\boxpi) &= {{\a^2} \over 2}({{{\tilde\om}}} +
 {D \over {x^3}}) {\tilde f}\cr
{{{\tilde\om}}} &= -2(\pip)^2 (\pip + {3\over x}) \cr
{\tilde f} &= \pip (\pip +{2\over x})} \eqno(12) $$

To preserve the Pontryagin index and finiteness of energy the
boundary conditions on $\Pi$ are taken to be the same as before i.e.
$\Pi\lnrt {\r^2\over{x^2}}$ as $x\lnrt 0$ and
$\Pi\lnrt 1 +{\r^2\over{x^2}} $ as $x\lnrt\infty$

At first sight the extreme non linearity of these equations seems
intractable. One expects,however, that if one
can solve it in the asymptotic regions $x\lnrt 0,x\lnrt \infty$
to obtain a solution with the same winding number as for $N_f=0$ then
a solution which interpolates between
these regions may be obtained numerically. Evaluating the r.h.s of eqn(12)
in the limit $x\lnrt 0$ one
finds that the leading term is $-2\a^2(D-8)/(x^3\r^2)$  while the l.h.s is less
singular. Hence the choice $D =8$ is confirmed. With $D = 8$ one finds
that one can solve for the unknown coefficients in the deformed superpotential

 $$\Pi_0= 1 + {{\r^2}\over {x^2}} + a_1 x^2 + a_2 x^4 + \ldots \eqno(13)$$

to get

 $$ a_1 =-{{\a^2}\over {\r^4}}\qquad\qquad
 a_2 =+{{2\a^2}\over {3\r^6}}(1 +{{9\a^2}\over {4\r^2}})\eqno(14)$$

and so on. Note that inspite of the nonlinearity of the fermion back
reaction represented by the Chern-Simons term, the deformations are
entirely non singular. Similarly in the $x\rightarrow\infty$ region

  $$\Pi_{\infty}= 1 + {{\r^2}\over {x^2}} +
  {{ b_1}\over{ x^4}} +   {{ b_2}\over{ x^6}}  + \ldots \eqno(15)$$

 $$ b_1 ={{\a^2 {\r^2}}\over 3}\qquad\qquad
b_2 ={{\a^4 {\r^2}}\over {18}}\eqno(16)$$

A numerical integration of the equation in the intermediate region will
be given elsewhere. The asymptotic behaviour of the electric field is

$$\eqalign{ E_{\m}^8 &=-\partial_{\m} A_0^8 =
 -{\a \over 2}({{{\tilde{\om}}_{\m}}} +
 {{D {{{\hat{x}}_{\m}}}\over {x^3}}})\cr
 &= -{{4\a}\over {x^3}}{{{\hat{x}}_{\m}}} +
  {{12\a\r^4}\over{x^7 }} {{{\hat{x}}_{\m}}} +O(x^{-9})}\eqno(17)$$

The electric charge is

$$Q^8=\QNK\int_{S_{\infty}} d\Sigma_{\m} E_{\m}^8 =
-{{N_f}\over{4{\sqrt 3}}}\eqno(18)$$

The electric field near the origin is

$$E_{\m}^8= -{{12\a}\over {\r^4}}{{x_{\m}}} +O(x^3)\eqno(19)$$

The magnetic field is

$$\eqalign{\QNE &=-{4\over{\r^2}}(\QNII +
 2  {{{\hat{x}}_{\s}}}{{\bar\eta}_{a\s{[\m}}}{{{\hat{x}}_{\n ]}}})  + O(x^2)
 \qquad\qquad x\lnrt 0\cr
&=-{{4\r^2}\over{x^4}}(\QNII +
 2  {{{\hat{x}}_{\s}}}{{\bar\eta}_{a\s{[\m}}}{{{\hat{x}}_{\n ]}}})  + O(x^{-6})
 \qquad\qquad x\lnrt \infty}
 \eqno(20)$$

The magnetic field is thus asymptotically {\it{dipolar}}.

The CS term does not contribute to the expression for the field
energy :

$$\eqalign{{\cal E} &= {{8\p\n}\over{g_5^2}} +
 {3\over{2 g_5^2}}\int d^4x(\boxpi)^2
-{1\over{2g_5^2}}\int d^4x d^4y {\r}_8(x) G_0(x,y){\r}_8(y)\cr
{\r}_8(x) &=-{\a\over 4} \QNE\QNEB
\qquad\qquad G_0(x,y)=-{1\over{4\pi^2(x-y)^2}}}
\eqno(21)$$

Thus it is the sum of the original uncharged soliton energy plus a
positive definite contribution from the electrostatic energy and
another due to the deviation from self duality. The winding number is
unaffected by the weak deformations exhibited in equations (13)-(16).

$$\eqalign{\n &= {1\over{32\p^2}}\int d^4x \QNE \QNEB =
{1\over{16\p^2}}\int d^4x \partial_{\m}{\tilo}\cr
&={1\over 8} x^3\xmhat\tilo\mid^{\infty}_0=1 }\eqno(22)$$

This can be checked by transforming to a nonsingular gauge using the
usual${}^{12}$ transformation
{\hbox{$U=(x^4+i{\vec{\tau}\cdot{\vec{x}}})/{\mid x\mid}$.}}
The above solutions are parametrized by an arbitrary scale parameter $\r$.
However the energy ${\cal E}$  is not independent of $\r$  .
 Thus one expects that the present solution, if stable at all,
  will relax to that value of $\r$ which minimizes the energy.
  To estimate this value one may
approximate $\Pi$ by $\Pi_{0}$ and $\Pi_{\infty}$ in the regions
$x\in [0,\r],x\in[\r,\infty]$  respectively to obtain

$${\cal E}= {{8\p^2}\over {g_5^2}}+
{{80\p^2g_5^6N_f^4}\over {27(32\p^2)^4 \r^4}}
+{{cg_5^2N_f^2}\over {\r^2}}\eqno(23)$$

Where the last term is the obvious dimensional estimate for the
electrostatic energy in 4 space dimensions and c is a numerical constant.
In the approximation we have used , it appears that our solution is
unstable against growth of the free parameter $\r$ which will tend
to \ $\infty$ so as to saturate the the self-duality lower bound on the energy.
Note however that the corrections to the neutral soliton energy are $O(g_5^2)$
and $O(g_5^6)$ thus it quite possible that quantum corrections may
stabilize $\r$.

Since the electric and magnetic fields are orthogonal in the internal
space, the Poynting vector, and therefore also the usual
 (Belinfante-Bessel-Hagen) expression for the the angular
 momentum of the field configuration, is zero.

It is easy to see that solutions with winding number $K$ can be
approximated by superpositions of $\Pi_{0}$ 's centered at different
$x_{i}$ near those $x_{i}$ and by $1+\sum_i{\r_i^2}/x^2+ O(x^{-4})$
in the $x\rightarrow\infty$ region.
To generalize from $SU(3)$ to the $SU(N)$ case one simply replaces
\hbox{${\l^8={1\over{\sqrt 3}}diag(1,1,-2)}$} by \hfil\break
\hbox{$\{\l^{k^2-1} = {\sqrt{2\over {k(k-1)}}}
 diag(1_{k-1},1-k,0_{N-k});k=3,...,N\}$}
etc${}^{13}$.

In conclusion we have demonstrated the existence of charged
topological solitons in 5 dimensional YMCS theory. This extends
results on charged vortices in 3 demensional YMCS theory and supports
the conjecture that similar ``charging" by the CS term occurs as well
in higher odd dimensions. The generalizations of our solution to
$M^{4}\times S^{1}$ will be relevant to fleshing out at least the
simplest example of the ``syncyclonic scenario"${}^1$. The
semiclassical quantization of of these objects is therefore
studied in Ref.[9].

\noindent {\bf Acknowledgements:} I am grateful to A.Khare, V.Soni,
S.Wadia and R.Shankar for very useful discussions. I thank S.Kar
for drawing my attention to M.Kobayashi's paper${}^2$.

\noindent{\bf Note Added :} We[13] have now shown that the different
embeddings of $SU(2)\times U(1)^{N-2}$ in $SU(N)$ correspond to a
${}^NC_2$-plet of
topological solitons. Moreover the ``dibaryon embedding'' i.e
$SO(3)\times U(1)^{N-3}$ in $SU(N)$ gives
{\it{neutral}} solutions for $N=3$ and a ${}^NC_3$-plet for $N\geq 4$
(after modding out the action of certain $SU(3)$s).
\vfil
\eject

\centerline {\bf{References }}

\item {1)} C.S.Aulakh, {\it{Syncyclons or Solitonic Signals from Extra
Dimensions}},
\item{} IP/BBSR/92-14, to be published in Mod. Phys. Lett {\bf{A}}.
 See also Ref.[2].
\item {2)} A.Strominger Nucl. Phys. B 343 (1990) 167.
\item{} M.Kobayashi, Prog. Theor. Phys. 74(1985)1139.
\item {3)} B.J.Harrington and H.K.Shepard, Phys. Rev. D17 (1978), 2122.
\item {4)} D.Gross, R.Pisarski, L.Yaffe, Rev. Mod. Phys. 53 (1981) 43.
\item {5)} A.Niemi and G.W.Semenoff, Physics Reports 135 (1986) 99
and references therein ;
\item{}Phys. Rev. Lett. 51 (1983) 2077.
\item {6)} E.Witten,Phys.Lett. B86(1979)283.
\item {7)} E.Witten, Nucl. Phys. B223 (1983) 422.
\item{} S.Jain and S.R.Wadia , Nucl. Phys. B258(1985) 713.
\item {8)} S.K.Paul and A.Khare Phys. Lett. B174(1986)420,182(1986),414.
\item {} C.N.Kumar and A.Khare ,Phys.Lett. B 178(1986)385.
\item{} H.J.De Vega and F.A.Schaposhnik ,Phys. Rev. Lett. 56(1986)2564.
\item {9)} C.S.Aulakh and V.Soni, {\it{Collective Quantization of Topological
Solitons in 4+1 Dimensional YMCS Theory}}, IP/BBSR/92-37, to appear.
\item {10)} C.S.Aulakh, to appear.
\item {11)} G.'t Hooft (unpublished);Phys. Rev. D14(1976)3432.
\item {} E.Corrigan and D.Fairlie, Phys. Lett. B67 (1977), 69.
\item {} R.Jackiw, C.Nohl. and C.Rebbi, Phys. Rev. D15, (1977)1642.
\item {12)} R.Rajaraman, {\it Solitons and Instantons}, North Holland,
1982 and references therein.
\item {13)} C.S.Aulakh and V.Soni, {\it{ Topological
Soliton Multiplets in 4+1 Dimensional YMCS Theory}}, IP/BBSR/92-36, to appear.

\vfil
\eject
\end